\documentclass[twocolumn,secnumarabic,amssymb, nobibnotes, aps, prl,floatfix,superscriptaddress]{revtex4-1}
\usepackage{amsmath}
\usepackage{amssymb}

\usepackage{comment}

\usepackage{graphicx} 
\usepackage{hyperref}
\usepackage{babel}
\begin{document}
\title{Superradiant Neutrino Lasers from Radioactive Condensates}
\author{B.J.P. Jones}
\thanks{\url{ben.jones@uta.edu}}
\affiliation{Department of Physics, University of Texas at Arlington,
Arlington, Texas 76019, USA}
\author{J. A. Formaggio}
\thanks{\url{josephf@mit.edu}}
\affiliation{Laboratory for Nuclear Science and Dept. of Physics,
Massachusetts Institute of Technology, Cambridge, MA 02139, USA}
\begin{abstract}
Superradiance emerges from collective spontaneous emission in optically pumped gases, and is characterized  by photon emission enhancements of up to $\frac{1}{4}N^{2}$ in an $N$ atom system.  The gain mechanism derives from correlations developed within the decay medium rather than from stimulated emission as in lasing, so an analog of this process should be possible for fermionic final states.  We introduce here the concept of superradiant neutrino emission from a radioactive Bose Einstein condensate, which can form the basis for a superradiant neutrino laser.  A plausible experimental realization based on a condensate of electron-capture isotope $^{83}$Rb could exhibit effective radioactive decay rates accelerated from 86.2 days to minutes in viably sized rubidium condensates of $10^{6}$ atoms.
\end{abstract}
\maketitle

\noindent The neutrino is a nearly
massless particle whose feeble environmental interactions permit coherent
quantum mechanical effects. The interferences between neutrino mass states with different de Brogile wavelengths (oscillations)~\cite{giunti2007fundamentals}, and their modification by refraction in matter~\cite{kuo1989neutrino,mikheyev1989resonant}, are prominent and widely studied examples.  Within the high neutrino density of supernovae, entanglement through elastic scattering
is predicted to lead to collective neutrino oscillations
\cite{duan2010collective}; and the large cross-section enhancement associated with coherent elastic neutrino
nucleus scattering \cite{akimov2017observation} is a result of in-phase summation of scattering amplitudes from many nucleons. 

While some of these coherent phenomena have direct analogies in optics~\cite{weiner2017phase}, the interference effects available to neutrinos and photons are inequivalent due to the differences between their fermionic and bosonic statistics.   One form of coherence that is available to photons but not to neutrinos is lasing.  Lasers rely on  coherent amplification by stimulated emission (SE), only admissible for bosons.
Stimulation of neutrino pairs by
photons is possible, and has been invoked as a way to
measure the mass of the neutrino \cite{song2016conditions,zhang2016improved,yoshimura2014radiative,yoshimura2007neutrino,ge2023unique}, but 
direct laser amplification of neutrinos is prohibited because the allowed occupancy of any fermionic mode is limited to zero or one.

A different coherent amplification effect that has been observed with photons
is superradiance (SR) \cite{rehler1971superradiance}. Originally proposed
by Dicke \cite{dicke1954coherence}, in SR the collective
spontaneous emission of $N > 1$ atoms separated by less than one wavelength
of light contribute coherently to enhance the radiative decay rates by a  factor of up to $\frac{1}{4}N^2$. Following the
first observation in optically pumped HF gas \cite{skribanowitz1973observation},
SR has also been observed in several other contexts. 
Unlike in lasing, whereby each photon reinteracts with another atom in a
cavity to induce SE, derivations of SR stipulate that ``any photon, once produced, escapes from the active volume so
fast that it does not have a chance to feed itself back into atomic
excitation''~\cite{bonifacio1971quantum}. Devices exploiting
SR as their primary optical gain mechanism have been realized
\cite{haake1993superradiant,weiner2017phase}, and dubbed ``superradiant lasers'' (admittedly a misnomer, given the absence of SE in these devices).
 Ref.~\cite{bohnet2012steady} operates such a device with $\leq1$ photon
in the cavity at a time, a strong demonstration that SE is not involved. SR lasers store optical coherence and phase correlations in the gain medium rather than the electromagnetic field, and enhancement occurs through collective spontaneous emission~\cite{akkermans2008photon,bonifacio1971quantum,gross1982superradiance,haake1993superradiant,rehler1971superradiance,schuurmans1982superfluorescence}. This mechanism does not rely intrinsically on the bosonic nature of light and so should also be admissible for fermions.

In this Letter we discuss application of  SR  to produce coherent, amplified neutrino sources.  Decays by electron capture (EC)  provide a particularly close analogy to the model two-state systems previously invoked in discussions of SR~\cite{dicke1954coherence,gross1982superradiance}. SR $\beta$ decay also appears to be viable, with cold atomic T being one possible candidate isotope. Finally, the superradiant state also exhibits amplified neutrino interaction rates through enhanced collective neutrino absorption. 

One challenge for realizing SR with neutrinos is that the lowest energy detectable neutrinos are produced in $\beta$ or $EC$ decays with energies of at least a few keV.  A requirement for  ordinary SR is that
the atom originating any photon (neutrino) should not be distinguishable
based measurements that can be made upon that photon (neutrino),  typically interpreted as a statement
that the emitters must be closer together than the DeBrogile wavelength $\lambda$.
This is one way to achieve such indiscernibility, but there are others. For example, if all
initial atoms are prepared in an identical spatial wave function,
as in a Bose Einstein Condensate (BEC)\cite{griffin1996bose,pethick2008bose,schneble2003onset}, coherent spontaneous radioactive emission of the daughter particles should also be enabled.  This is a proposed method for formation of a gamma ray laser~\cite{avetissian2014self}, and related forms of collective emission from a BEC have been discussed as mechanisms to achieve implementations based on trapped  positronium- ~\cite{avetissian2015gamma} gamma-decaying nuclear isomers~\cite{marmugi2018coherent}.  Here we investigate the applicability of SR from a BEC to coherently amplified neutrino production.


\section{Enhancement Mechanism}
The phenomenon
of Dicke SR \cite{rehler1971superradiance} or superfluorescence
\cite{schuurmans1982superfluorescence} occurs through the accumulation of quantum
correlations between decayed and undecayed atoms in an ensemble. Consider a collection of $N$ identical two-level
atoms with upper and lower states labeled $|g_i\rangle$
and $|e_i\rangle$. We can define raising and lowering operators
for atom $i$, as well as the $D_{i}^{3}$ operator, 
\begin{eqnarray}
D_{i}^{+}&=&|e_{i}\rangle\langle g_{i}|.\quad D_{i}^{-}=|g_{i}\rangle\langle e_{i}|,\\
D_{i}^{3}&=&\frac{1}{2}\left[|e_{i}\rangle\langle e_{i}|-|g_{i}\rangle\langle g_{i}|\right],
\end{eqnarray}
The $D$ operators have commutation rules,
\begin{equation}
[D_{i}^{3},D_{j}^{\pm}]=\pm\delta_{ij}D_{i}^{\pm},\quad[D_{i}^{+},D_{j}^{-}]=2\delta_{ij}D_{i}^{3},
\end{equation}
and the rate of decay to a photon is driven by interaction Hamiltonian $H_{int}$, which under the rotating wave approximation (RWA) is,
\begin{equation}
H_{int,i}=-id_{a}E_{l}(D_{i}^{+}\hat{a_{l}}+D_{i}^{-}\hat{a_{l}}^{\dagger}).
\end{equation}
$E_{l}$ is the electric field strength associated with one photon
and $d_{a}=\langle e|\hat{x}|g\rangle$ is the dipole matrix element that
controls the rate of optical transitions. In a sufficiently optically pumped gas
we can consider the initial state to be fully excited, $|\psi(0)\rangle=|e_1e_2...e_N\rangle$. Despite all atoms being in $|e\rangle$ we emphasize that this is {\em not} a BEC, since we are labeling
only the excitation level of each atom; each also posses a position-space wave function, which for traditional SR can be assumed to be distinct and non-overlapping.

The simplest SR condition 
is that all atoms are motionless and contained in a small volume compared to
the transition wavelength $\lambda=hc/(E_e-E_g)$. In this case we can express the full interaction Hamiltonian as
\begin{equation}
H_{int}=\sum_{i}H_{int,i}=-id_{a}E_{l}(D^{+}\hat{a_{l}}+D^{-}\hat{a_{l}}^{\dagger}),
\end{equation}
where we have identified the ``D-spin'' of the entire system $D\equiv\sum_{i}D_{i}$.  The eigenstates of $D$ are constructed by analogy to coupling many spin-1/2
angular momenta. The largest $J$-plet is comprised of $N$ states with
quantum number $M$ ranging between $\pm\frac{1}{2}N$, obtained via
the lowering operation on $|\psi(0)\rangle$
\begin{equation}
|JM\rangle=\sqrt{\frac{\left(J+M\right)!}{N!(J-M)!}}\left(D^{-}\right)^{J-M}|eeee...e\rangle.
\end{equation}

The $|JM\rangle$ state has $J+M$ atoms in $|e\rangle$ and $J-M$
atoms in $|g\rangle$, appropriately symmetrized. Photon
emission is generated by the $D^{-}$ in $H_{int}$, so each emission ladders downwards between
elements of the same $J$-multiplet.  The rate of emission $W_N$  is found via Fermi's golden rule, 
\begin{equation}
W_N\propto|\langle J,M-1|H_{int}|J,M\rangle|^{2}.\label{eq:FGR}
\end{equation}
Only one final state has a non-zero amplitude given this initial one, so Eq.~\ref{eq:FGR} can be written
\begin{eqnarray}
W_N&=&\sum_{f}\langle JM|D^{+}\hat{a}|f\rangle\langle f|D^{-}\hat{a}^{\dagger}|JM\rangle\Gamma_\gamma\\
&=&\langle D^{+}D^{-}\left(1+\hat{a}^{\dagger}\hat{a}\right)\rangle\Gamma_\gamma,
\end{eqnarray}
where $\Gamma_\gamma$ is the single particle rate, $[\hat{a},\hat{a}^{\dagger}]=1$ has been employed to move $\hat{a}$ to the  right, and the expectation $\langle\rangle$ is implicitly taken for the initial state only. For emission into a cavity that already contains $n$ photons we find a decay rate enhanced by $1+n$,
\begin{equation}
W_N=\left(1+n\right)\langle D^{+}D^{-}\rangle\Gamma_\gamma,
\end{equation}
demonstrating conventional lasing through SE. When $n=0$ photons are present the rate is instead
\begin{eqnarray}
W_{N}&=&\langle D^{+}D^{-}\rangle\Gamma_\gamma\\
&=&(J+M)(J-M+1)\Gamma_\gamma.
\end{eqnarray}
$W_N$ increases from
$N\Gamma_\gamma$ when $M=J=\frac{1}{2}N$ (nothing has decayed), to $\frac{1}{2}N(\frac{1}{2}N+1)\Gamma_\gamma$
when $M=0$ (half of the atoms have decayed).  For large $N$, a dramatic enhancement for the partially-decayed state beyond the ordinary fluorescence rate $(J+M)\Gamma_\gamma$ is observed, embodying  SR amplification.

Consider now the analogous system involving a BEC that undergoes $EC$ to produce neutrinos. The two atomic basis states are a parent $|p_{i}\rangle$ and a daughter emitted alongside a neutrino $|d_{i}\rangle$. For both $\beta$ decay and $EC$, the analogous operators to $D^{\pm}$ are isospin ladders $\tau^{\pm}$. The total interaction Hamiltonian can be written, this time without requiring the RWA, as
\begin{equation}
H^{\nu}_{int}=\sum_{i}(\tau_{i}^{+}\hat{a}_{\nu}+\tau_{i}^{-}\hat{a}_{\nu}^{\dagger}).
\end{equation}
In a BEC all
neutrinos necessarily emerge from indistinguishable atomic states, so we no longer require the long wavelength assumption. Equivalent arguments are then used
obtain the total emission rate $W^{\nu}_{N}$,
\begin{equation}
W^{\nu}_{N}=\langle{\cal T}^{+}{\cal T}^{-}\left(1-\hat{a}_{\nu}^{\dagger}\hat{a}_{\nu}\right)\rangle\Gamma_\nu,
\end{equation}
now in terms of ${\cal T}=\sum_i \tau_i$ and applying fermionic commutation relations $\{\hat{a}_{\nu},\hat{a}_{\nu}^{\dagger}\}=1$. 

Once again we distinguish between the cases where there is / is not a neutrino
in the emission region at the time of decay. With a neutrino present, $\hat{a}_{\nu}^{\dagger}\hat{a}_{\nu}|i\rangle=|i\rangle$
and so $W^{\nu}_N=0$: Pauli blocking inhibits the lasing effect that was observable with photons. On the other hand, if each
 neutrino quickly leaves then  $\hat{a}_{\nu}^{\dagger}\hat{a}_{\nu}|i\rangle=0$, again yielding
\begin{equation}
W^{\nu}_{N}=(J+M)(J-M+1)\Gamma_\nu.
\end{equation} 
Thus SR emission is expected for a radioactive BEC decaying to neutrinos.

The quantum correlations leading to SR accumulate within the gain
medium during decay.  In the BEC example,
the decay daughters are likely to escape and interact with the environment
on timescales faster than this accumulation occurs. We must therefore
 examine whether such interactions destroy 
coherent amplification process. To this end, we consider the process again incorporating both the state label ($|e\rangle$ vs $|g\rangle$ or $|p\rangle$ vs $|d\rangle$) and also the position wave function of each atom $|\phi_{i}\rangle$.
Under the original Dicke mechanism, the $\phi_{i}$ are distinguishable and non-overlapping.

Considering a three-atom system for illustration, the undecayed state $|\psi_{0}\rangle$ decays to produce the first photon as
$|\psi_{0}\rangle\rightarrow |\psi'_{0}\rangle$ with
 rate \textbf{$W_{N=3}^{1\gamma}$},
\begin{eqnarray}
|\psi_{0}\rangle&=&|\frac{3}{2},\frac{3}{2}\rangle=|e_{1}e_{2}e_{3}\rangle\otimes|\phi_{1}^{e}\phi_{2}^{e}\phi_{3}^{e}\rangle\\
\rightarrow|\psi'_{0}\rangle&=&\left(|e_{1}e_{2}g_{3}\rangle\otimes|\phi_{1}^{e}\phi_{2}^{e}\phi_{3}^{g}\rangle+\{...\}\right)\otimes|\gamma\rangle,\quad\quad \\
W_{N=3}^{1\gamma}&=&\sum_{\lambda}|\langle\lambda|\psi_{1}\rangle|^{2}\Gamma_{\gamma}=3\Gamma_{\gamma}.
\end{eqnarray}
The $\{...\}$ here denotes summation over permutations of 1,2,3, and
$\sum_\lambda$ implies both summation over flavor and integration over position bases. For the first decay $W_N=N\Gamma$ so there is no SR enhancement, as expected.
More interesting is the second decay, which occurs from a
Dicke state,
\begin{eqnarray}
|\psi_{1}\rangle&=&\frac{|e_{1}e_{2}g_{3}\rangle\otimes|\phi_{1}^{e}\phi_{2}^{e}\phi_{3}^{g}\rangle+\{...\}}{\sqrt{3}}\\
\rightarrow|\psi'_{1}\rangle&=&\frac{|g_{1}e_{2}g_{3}\rangle\otimes|\phi_{1}^{g}\phi_{2}^{e}\phi_{3}^{g}\rangle+\{...\}}{\sqrt{3}}\otimes|\gamma\rangle.\quad\quad\label{eq:SecondDecay}\\
W_{N=3}^{2\gamma}&=&\sum_{\lambda}|\langle\lambda|\psi\rangle|^{2}\Gamma_{\gamma}=4\Gamma_{\gamma},
\end{eqnarray}
Only two $|e\rangle$ are
present, yet the decay rate scales as $4\Gamma_\gamma$ due to SR. The origin of the enhancement is that, for example, the $|g_{1}e_{2}g_{3}\rangle$ in $|\psi'_{1}\rangle$
is generated indistinguishably from both the  $|e_{1}e_{2}g_{3}\rangle$ and $|g_{1}e_{2}e_{3}\rangle$ terms
in $|\psi_{1}\rangle$, which add coherently.  This interpretation is further developed in~\cite{wiegner2011quantum}, which also explores the effects of relaxing the long-wavelength approximation. 

Decohering interactions can be included in this model via a Von-Neumann
measurement, in which the environment prepared in initial state $|\epsilon_{0}\rangle$
becomes entangled with one of the daughter particles $i$
via 
\begin{equation}
|\phi_{i}^{g}\rangle\otimes|\epsilon_{0}\rangle\rightarrow|\phi'_{i}\rangle\otimes|\epsilon_{i}\rangle,
\end{equation} 
and then with another atom $j$, as 
\begin{equation}
|\phi'_{i}\rangle\otimes|\phi_{j}^{g}\rangle\otimes|\epsilon_{i}\rangle\rightarrow|\phi'_{i}\rangle\otimes|\phi'_{j}\rangle\otimes|\epsilon_{ij}\rangle.
\end{equation}
We assume that the environment is not so altered by $i$ that $|\phi'_{j}\rangle$ depends strongly on whether $j$ or $i$ was emitted first; but that the final state of environment is nevertheless distinct for the two orderings, $|\epsilon_{ij}\rangle\neq|\epsilon_{ji}\rangle$. Under these assumptions a modified decay rate is derived,
\begin{equation}
\tilde{W}_{N=3}^{2\gamma}=\frac{2}{3}\left(3+\sum_{i<j}\mathcal{R}\langle\epsilon_{ij}|\epsilon_{ji}\rangle\right)\Gamma_\gamma.\label{eq:DecRate}
\end{equation}
As is typical in decoherence models, Eq.~\ref{eq:DecRate} has a coherent limit when the environment contains no distinguishing information about the emission order,  $\langle\epsilon_{ij}|\epsilon_{ji}\rangle=1$ and $W_{N=3}^{2\gamma}=4\Gamma_\gamma$; and an incoherent one when the different entangled environment states are completely 
distinguishable,  $\langle\epsilon_{ij}|\epsilon_{ji}\rangle=\delta_{ij}$ and $W_{N=3}^{2\gamma}=2\Gamma_\gamma$. These limits correspond to superradiant emission and ordinary fluorescence, respectively. 

For neutrino emission from a radioactive BEC, the
spatial wave functions are necessarily all identical, $|\phi_{i}^{p}\rangle=|\phi^{p}\rangle$, and $|\phi_{i}^{d}\rangle=|\phi^{d}\rangle$.
As a result, environmental interactions of the daughters all take the same form 
\begin{equation}
|\phi^{d}\rangle\otimes|\epsilon_{0}\rangle\rightarrow|\phi'\rangle\otimes|\epsilon\rangle,
\end{equation}
where in this case $|\epsilon\rangle$ has no dependence on $i$ at all. 
Since the environment state now carries no subscripts, coherent amplification is always maintained,
\begin{equation}
\tilde{W}_{N=3}^{2\nu}=4\Gamma_{\nu},
\end{equation}
It is remarkable that for emission from the fully spatially degenerate initial state provided by a BEC, coherence appears possible even in the presence of MeV-scale momentum transfers associated with the recoiling daughter atoms.  This is so because even though the recoil can transfer information about its origin to the surroundings, it remains impossible to ascertain which parent atom has decayed. Furthermore, since it is the rate of neutrino-daughter pairs that is amplified, the entanglement between the neutrinos and the recoils also does not suppress coherence.  Rather, the correlations among trapped parent atoms within the decaying condensate is sufficient to preserve SR enhancement.

\begin{figure}[t]
\begin{centering}
\includegraphics[width=0.99\columnwidth]{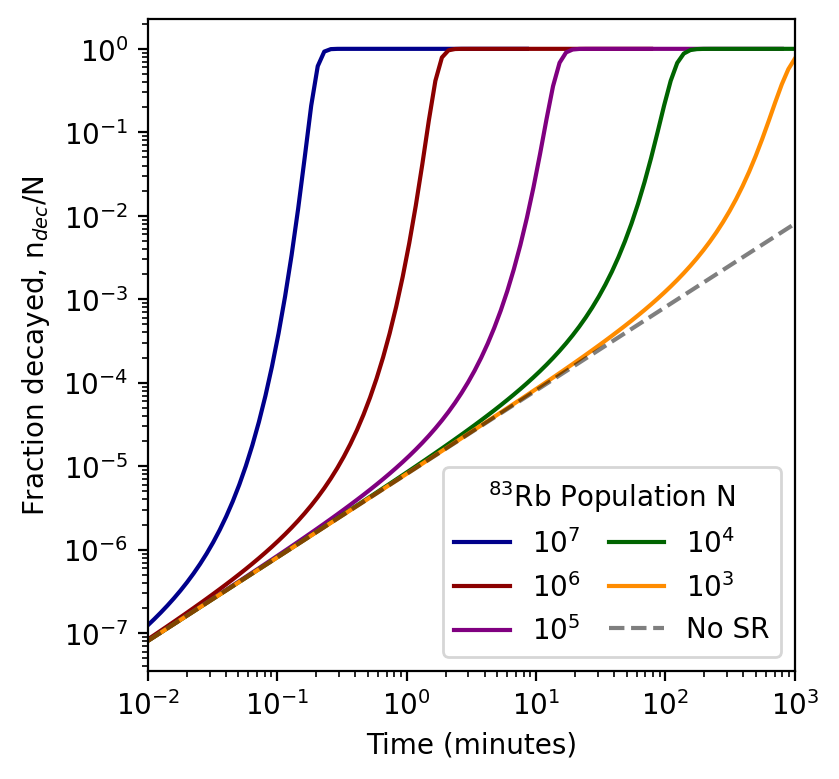}
\par\end{centering}
\caption{Comparison of the SR and ordinary fluorescence decay rates in $^{86}$Rb.\label{fig:DecayRate}}
\end{figure}

\section{Amplified $\nu$ emission}
To realize this effect experimentally, we require an isotope
that: 1) is radioactive; 2) has a bosonic neutral atom; 3) has sufficiently short half life that reaching a $\sim$50\% population  is viable during  a trappable lifetime; 4) is practically coolable to degeneracy, ideally with laser cooling; 5) offers a method for monitoring the decay rate. 

Several candidates in the table of nuclides meet these criteria, one example being $^{83}\mathrm{Rb}$. The neutral atom contains 46 neutrons, 37
protons and 37 electrons, meaning it is a boson and can be condensed. It decays 
via EC to $^{83m}\mathrm{Kr}$, which can be counted
by subsequent decay X-rays produced by de-excitation of the decay daughter with half-life of 1.83 hours. $^{83}\mathrm{Rb}$ has
a half-life of $\tau_{1/2}=86.2$ days,  which is long enough for convenient
handling but short enough for depletion to be measured on viable timescales. 

Other isotopes of Rb are frequently  laser cooled to produce BEC,
with works such as Ref \cite{lin2009rapid} producing  $^{87}$Rb condensates
with $2\times10^{6}$ atoms at a 16~s repetition rate.  As in most studies of rubidium BECs, Ref \cite{lin2009rapid} employs adiabatic cooling, using a 49~$\mu$K magnetic trap loaded through a sequence of optical pre-cooling methods. This leads to a final condensate temperature of of 0.32\, $\mu$K, comfortably below the critical temperature of $T_c=0.5\, \mu$K.  Condensates of $^{85}\mathrm{Rb}$ have also been produced~\cite{otoishi2020rapid}. There, polarization gradient cooling in a 3D optical lattice and a compressible double-crossed dipole trap were employed, in order to circumvent difficulties associated with the more difficult collisional properties of $^{85}$Rb relative to $^{87}$Rb.  Condensates of $N\sim 2\times 10^5$ $^{85}$Rb atoms were realized, on a few-second timescales. No condensates of $^{83}$Rb have been reported to date, though its collisional properties have been presented in Ref.~\cite{burke1999ultracold} as being remarkably similar to those of $^{87}$Rb.  Ref.~\cite{burke1999ultracold} also discusses the possible production of a double condensate of $^{87}$Rb and $^{83}$Rb, enabled through a convenient and tunable Feshbach resonance. In such a system the non-radioactive $^{87}$Rb would serve as a passive background species and not be expected to interfere with the emission phenomena, and so this mixture could also serve as a convenient SR neutrino laser gain medium as an alternative to an ensemble of pure $^{83}$Rb.

Assuming a suitable BEC can be produced, the effects of
SR enhancement become apparent
quickly. The number of decayed atoms $n_{dec}$ exhibiting exponential growth,
\begin{equation}
n_{dec}(t) = N \frac{e^{(N+1)\Gamma_\nu t}-1}{e^{(N+1)\Gamma_\nu t}+N}.
\end{equation}
We quantify the gain associated with the SR process by constructing an equivalent ``half-life" $(\tilde{\tau}_{1/2})$.  In the limit that $N\gg 1$, finding
\begin{equation}
    \frac{\tilde{\tau}_{1/2}}{\tau_{1/2}} \simeq \frac{\ln{(N)}}{(N+1)\ln{(2)}}.
\end{equation}
For $^{83}\mathrm{Rb}$ the accelerated half-life drops to $\tilde{\tau}_{1/2} \simeq 148$~s with a BEC of 10$^6$ atoms.  Fig.~\ref{fig:DecayRate} shows the population evolution under the SR mechanism (solid lines) in comparison to the ordinary radioactive decay (dashed) for various condensate sizes, illustrating this large enhancement. 

\section{Amplified $\nu$ Capture}
Discussion of amplified decay rates invites consideration of amplified neutrino capture on the SR state.  Aspects of this problem have recently been discussed by others, in the context of axion, dark matter, and reactor neutrino searches~\cite{arvanitaki2024superradiant}.  This could have various applications, including   detection of the relic neutrino density from the primordial Big Bang.  Tritium (T) is one isotope commonly  targeted for relic neutrino capture searches, via $\nu_e + ~^3{\rm H} ~\rightarrow~ e^- + ~^3{\rm He}$, and cooling of T is an active ongoing effort for direct neutrino mass searches~\cite{esfahani2022project}.  The rate of neutrino capture on T is determined by the overlap of the BEC and neutrino spatial wavefunctions, and is directly related to the $\beta$ decay rate of T.  Cocco {\em et al}~\cite{AGCocco_2008} provide an estimate of connecting the half-life of beta decay to neutrino capture,
\begin{eqnarray}
\lambda_\nu & \simeq & \int \frac{G^2_F}{\pi}  \frac{p_e E_e F(Z,E_e)}{\exp{(p_\nu/T_\nu)}+1}\frac{d^3 p_\nu}{(2\pi)^3} \\
 & \simeq & 0.66\times 10^{-23} \Gamma_\beta
\end{eqnarray}
\noindent where $\Gamma_\beta$ is the tritium beta decay rate ($2\times10^{-9}$ s$^{-1}$). 

The SR mechanism does give an enhancement to this process which is analogous to the enhancement provided for emission from the Dicke state. However, even in this accelerated mode one requires of order $10^{20-25}$ atoms and hold times of order a year to detect a few relic neutrino events.  This is mainly because one needs to initiate the decay of a few events before the enhancement process can start to take hold.  As such, while the SR mechanism indeed appears to have implications for accelerated  neutrino capture, it does not appear to offer obvious sensitivity to the cosmological relic neutrino background.

\section{Conclusion}
We have introduced the concept of a superradiant source for enhanced neutrino emission via collective spontaneous decay of a radioactive BEC.  The isotope $^{83}$Rb is presented as a promising candidate, with an 86 day half-life that may be accelerated to minutes by superradiance. 

Many aspects are absent from our discussion. These include a detailed presentation of a method for cooling the radioactive vapor to degeneracy; management of condensate heating by escaping particles and their energy losses into the trap; the method of decay rate measurement; and others.  We defer such practical considerations to future work.   Our primary conclusion is that there is a mechanism by which coherent, enhanced, ``laser-like'' emission of neutrinos is possible, and that it is potentially experimentally observable.  The  enhancement factor is controllable via the size of condensate population, and increases of order 10$^6$ are accessible with reasonably sized, radioactive Bose Einstein condensates.  

\section{acknowledgements}
BJPJ is supported by the US Department of Energy under DOE-SC0019223 and DE-SC0024434.  JAF is supported by the US Department of Energy Award No.~DE-SC0011091 and the National Science Foundation Grant No.~PHY-2110569.  The authors wish to thank David Moore, Kyle Leach and Joachim Kopp for their helpful feedback and discussion.

\bibliographystyle{unsrt}
\bibliography{bilbio}

\begin{thebibliography}{10}

\bibitem{giunti2007fundamentals}
Carlo Giunti and Chung~W Kim.
\newblock {\em Fundamentals of neutrino physics and astrophysics}.
\newblock Oxford university press, 2007.

\bibitem{kuo1989neutrino}
Tzee-Ke Kuo and James Pantaleone.
\newblock Neutrino oscillations in matter.
\newblock {\em Reviews of Modern Physics}, 61(4):937, 1989.

\bibitem{mikheyev1989resonant}
SP~Mikheyev and A~Yu Smirnov.
\newblock Resonant neutrino oscillations in matter.
\newblock {\em Progress in Particle and Nuclear Physics}, 23:41--136, 1989.

\bibitem{duan2010collective}
Huaiyu Duan, George~M Fuller, and Yong-Zhong Qian.
\newblock Collective neutrino oscillations.
\newblock {\em Annual Review of Nuclear and Particle Science}, 60(1):569--594, 2010.

\bibitem{akimov2017observation}
D~Akimov, JB~Albert, P~An, C~Awe, PS~Barbeau, B~Becker, V~Belov, A~Brown, A~Bolozdynya, B~Cabrera-Palmer, et~al.
\newblock Observation of coherent elastic neutrino-nucleus scattering.
\newblock {\em Science}, 357(6356):1123--1126, 2017.

\bibitem{weiner2017phase}
Joshua~M Weiner, Kevin~C Cox, Justin~G Bohnet, and James~K Thompson.
\newblock Phase synchronization inside a superradiant laser.
\newblock {\em Physical Review A}, 95(3):033808, 2017.

\bibitem{song2016conditions}
Ningqiang Song, R~Boyero~Garcia, JJ~Gomez-Cadenas, Ma~Concepci{\'o}n Gonzalez-Garcia, A~Peralta~Conde, and Josep Taron.
\newblock Conditions for statistical determination of the neutrino mass spectrum in radiative emission of neutrino pairs in atoms.
\newblock {\em Physical Review D}, 93(1):013020, 2016.

\bibitem{zhang2016improved}
Jue Zhang and Shun Zhou.
\newblock Improved statistical determination of absolute neutrino masses via radiative emission of neutrino pairs from atoms.
\newblock {\em Physical Review D}, 93(11):113020, 2016.

\bibitem{yoshimura2014radiative}
M~Yoshimura and N~Sasao.
\newblock Radiative emission of neutrino pair from nucleus and inner core electrons in heavy atoms.
\newblock {\em Physical Review D}, 89(5):053013, 2014.

\bibitem{yoshimura2007neutrino}
M~Yoshimura.
\newblock Neutrino pair emission from excited atoms.
\newblock {\em Physical Review D - Particles, Fields, Gravitation, and Cosmology}, 75(11):113007, 2007.

\bibitem{ge2023unique}
Shao-Feng Ge and Pedro Pasquini.
\newblock Unique probe of neutrino electromagnetic moments with radiative pair emission.
\newblock {\em Physics Letters B}, 841:137911, 2023.

\bibitem{rehler1971superradiance}
Nicholas~E Rehler and Joseph~H Eberly.
\newblock Superradiance.
\newblock {\em Physical Review A}, 3(5):1735, 1971.

\bibitem{dicke1954coherence}
Robert~H Dicke.
\newblock Coherence in spontaneous radiation processes.
\newblock {\em Physical review}, 93(1):99, 1954.

\bibitem{skribanowitz1973observation}
N~Skribanowitz, IP~Herman, JC~MacGillivray, and MS~Feld.
\newblock Observation of dicke superradiance in optically pumped hf gas.
\newblock {\em Physical Review Letters}, 30(8):309, 1973.

\bibitem{bonifacio1971quantum}
R~Bonifacio, P~Schwendimann, and Fritz Haake.
\newblock Quantum statistical theory of superradiance. i.
\newblock {\em Physical Review A}, 4(1):302, 1971.

\bibitem{haake1993superradiant}
Fritz Haake, Mikhail~I Kolobov, Claude Fabre, Elisabeth Giacobino, and Serge Reynaud.
\newblock Superradiant laser.
\newblock {\em Physical review letters}, 71(7):995, 1993.

\bibitem{bohnet2012steady}
Justin~G Bohnet, Zilong Chen, Joshua~M Weiner, Dominic Meiser, Murray~J Holland, and James~K Thompson.
\newblock A steady-state superradiant laser with less than one intracavity photon.
\newblock {\em Nature}, 484(7392):78--81, 2012.

\bibitem{akkermans2008photon}
E~Akkermans, A~Gero, and R~Kaiser.
\newblock Photon localization and dicke superradiance in atomic gases.
\newblock {\em Physical review letters}, 101(10):103602, 2008.

\bibitem{gross1982superradiance}
Michel Gross and Serge Haroche.
\newblock Superradiance: An essay on the theory of collective spontaneous emission.
\newblock {\em Physics reports}, 93(5):301--396, 1982.

\bibitem{schuurmans1982superfluorescence}
MFH Schuurmans, QHF Vrehen, D~Polder, and HM~Gibbs.
\newblock Superfluorescence.
\newblock In {\em Advances in atomic and molecular physics}, volume~17, pages 167--228. Elsevier, 1982.

\bibitem{griffin1996bose}
Allan Griffin, David~W Snoke, and Sandro Stringari.
\newblock {\em Bose-einstein condensation}.
\newblock Cambridge University Press, 1996.

\bibitem{pethick2008bose}
Christopher~J Pethick and Henrik Smith.
\newblock {\em Bose--Einstein condensation in dilute gases}.
\newblock Cambridge university press, 2008.

\bibitem{schneble2003onset}
Dominik Schneble, Yoshio Torii, Micah Boyd, Erik~W Streed, David~E Pritchard, and Wolfgang Ketterle.
\newblock The onset of matter-wave amplification in a superradiant bose-einstein condensate.
\newblock {\em Science}, 300(5618):475--478, 2003.

\bibitem{avetissian2014self}
HK~Avetissian, AK~Avetissian, and GF~Mkrtchian.
\newblock Self-amplified gamma-ray laser on positronium atoms from a bose-einstein condensate.
\newblock {\em Physical review letters}, 113(2):023904, 2014.

\bibitem{avetissian2015gamma}
HK~Avetissian, AK~Avetissian, and GF~Mkrtchian.
\newblock Gamma-ray laser based on the collective decay of positronium atoms in a bose-einstein condensate.
\newblock {\em Physical Review A}, 92(2):023820, 2015.

\bibitem{marmugi2018coherent}
Luca Marmugi, Philip~M Walker, and Ferruccio Renzoni.
\newblock Coherent gamma photon generation in a bose--einstein condensate of 135mcs.
\newblock {\em Physics Letters B}, 777:281--285, 2018.

\bibitem{wiegner2011quantum}
R~Wiegner, J~Von~Zanthier, and Girish~S Agarwal.
\newblock Quantum-interference-initiated superradiant and subradiant emission from entangled atoms.
\newblock {\em Physical Review A—Atomic, Molecular, and Optical Physics}, 84(2):023805, 2011.

\bibitem{lin2009rapid}
Y-J Lin, Abigail~R Perry, Robert~L Compton, Ian~B Spielman, and James~V Porto.
\newblock Rapid production of r 87 b bose-einstein condensates in a combined magnetic and optical potential.
\newblock {\em Physical Review A, Atomic, Molecular, and Optical Physics}, 79(6):063631, 2009.

\bibitem{otoishi2020rapid}
Haruka Otoishi, Shu Nagata, Takumi Yukawa, Kazuya Yamashita, and Toshiya Kinoshita.
\newblock Rapid production of a rb 85 bose-einstein condensate in a double compressible optical dipole trap.
\newblock {\em Physical Review A}, 102(2):023316, 2020.

\bibitem{burke1999ultracold}
James~P Burke~Jr and John~L Bohn.
\newblock Ultracold scattering properties of the short-lived rb isotopes.
\newblock {\em Physical Review A}, 59(2):1303, 1999.

\bibitem{arvanitaki2024superradiant}
Asimina Arvanitaki, Savas Dimopoulos, and Marios Galanis.
\newblock Superradiant interactions of the cosmic neutrino background, axions, dark matter, and reactor neutrinos.
\newblock {\em arXiv preprint arXiv:2408.04021}, 2024.

\bibitem{esfahani2022project}
A~Ashtari Esfahani, S~B{\"o}ser, N~Buzinsky, MC~Carmona-Benitez, C~Claessens, L~De~Viveiros, PJ~Doe, S~Enomoto, M~Fertl, JA~Formaggio, et~al.
\newblock The project 8 neutrino mass experiment.
\newblock {\em arXiv preprint arXiv:2203.07349}, 2022.

\bibitem{AGCocco_2008}
A~G Cocco, G~Mangano, and M~Messina.
\newblock Probing low energy neutrino backgrounds with neutrino capture on beta decaying nuclei.
\newblock {\em Journal of Physics: Conference Series}, 110(8):082014, may 2008.

\end{thebibliography}

\end{document}